\documentclass[a4paper,twocolumn,showpacs,amsmath,pra,amssymb]{revtex4-1} 
\usepackage[T1]{fontenc}
\usepackage[latin9]{inputenc}
\usepackage{times}
\usepackage{color}
\usepackage{xspace}
\usepackage{color,ulem}
\usepackage{amssymb,amsmath}
\usepackage{amsbsy}
\usepackage{graphicx}
\usepackage{bm}
\usepackage{epstopdf}
\usepackage{float}

\usepackage[unicode,breaklinks]{hyperref}
\hypersetup{
    unicode=true,
    a4paper=true,
    plainpages=false, 
    colorlinks=true,
    linkcolor=blue,
    citecolor=blue,
    filecolor=black,
    urlcolor=blue
}
\newcommand{\diff}{\mathrm{d}}

\newcommand{\bx}{\ensuremath{\mathbf{x}}\xspace}
\newcommand{\x}{\mathbf{x}}

\newcommand{\bk}{\mathbf{k}}

\begin{document}
\newcommand{\nt}{\tilde{n}}
\newcommand{\Lnh}{\ensuremath{\mathcal{L}}\xspace}
\newcommand{\dbx}{\diff\bx}
\newcommand{\dbk}{\diff\bk}
\newcommand{\nbe}{\ensuremath{\bar{n}_{\mathrm{BE}}}\xspace}

\newcommand{\bro}{\mathbf{x}_1}
\newcommand{\brt}{\mathbf{x}_2}

\title{Roton excitations in a trapped dipolar Bose-Einstein condensate} 

\author{R.~N.~Bisset}  
\author{D.~Baillie}     
\author{P.~B.~Blakie}  
\email{blair.blakie@otago.ac.nz}
\affiliation{Jack Dodd Centre for Quantum Technology, Department of Physics, University of Otago, Dunedin, New Zealand.}

\begin{abstract}
We consider the quasiparticle excitations of a trapped dipolar Bose-Einstein condensate. By mapping these excitations onto linear and angular momentum we show that the roton modes are clearly revealed as discrete fingers in parameter space, whereas the other modes form a smooth surface. We examine the properties of the roton modes and characterize how they change with the dipole interaction strength. We demonstrate how  the application  of a perturbing potential can be used to engineer angular rotons, i.e.~allowing us to controllably select modes of non-zero projection of angular momentum to become the lowest energy rotons.
  \end{abstract}

\maketitle

\section{Introduction}
 
 Bose-Einstein condensates (BECs) with dipole-dipole interactions (DDIs) have been realized with highly magnetic atoms  \cite{Griesmaier2005a,*Pasquiou2011a,Mingwu2011a,*Lu2012a,Aikawa2012a}. This interaction is both long ranged and anisotropic and is predicted to open up an array of new phenomena for exploration using ultra-cold atomic gases \cite{Baranov2008,Lahaye_RepProgPhys_2009}. An important prediction is that a rotonlike excitation will emerge in a dipolar BEC which  is tightly confined along the direction that the dipoles are polarized \cite{Santos2003a}. There has been significant theoretical interest in  schemes for detecting rotons \cite{Corson2013a,Corson2013b,JonaLasinio2013b,Blakie2012a,Bisset2013a} and on the role of rotons in the behavior of dipolar BECs, such as response to perturbations \cite{Wilson2008a}, the critical velocity for the breakdown of superfluidity \cite{Wilson2010a,Ticknor2011a},  pattern formation \cite{Nath2010a,WIlson2012a}, and density fluctuations \cite{Bisset2013a,Blakie2013a,Klawunn2011}.

Initial theoretical predictions of Santos \textit{et al.}~\cite{Santos2003a} were made for a BEC of dipoles polarized and confined in the $z$ direction  (i.e.~untrapped in the $xy$-plane). In this case the quasiparticles are planewaves and the rotons occur as a local minimum in the dispersion relation  at wavevector  $k_{\mathrm{rot}}\sim1/a_z$, where $a_z$ is the $z$ confinement length \footnote{We also note that in tight two-dimensional geometry, where the DDI is purely repulsive, rotons can emerge under conditions of strong interactions or high density, however these occur at a wavevector set by the inverse particle spacing, e.g.~see \cite{Filinov2010a,Hufnagl2011a}.}.
While robust numerical techniques for calculating the quasiparticles of a fully trapped dipolar BEC have been developed (e.g.~see \cite{Ronen2006a}), there has been no comprehensive study of rotons for the trapped system. However, some aspects of the lowest energy rotons in the trapped system have emerged in studies of condensate structure and stability \cite{Ronen2007a,Asad-uz-Zaman2009a,Asad-uz-Zaman2010a,Asad-uz-Zaman2011a,Martin2012a,Ticknor2012b}. Recent work \cite{JonaLasinio2013} presented an approximate description of the trapped rotons by re-quantizing a local density treatment of the excitation spectrum, enabling an analytic prediction for the roton spectrum and wavefunctions.

\begin{figure} 
\begin{center}
\includegraphics[width=3.3in]{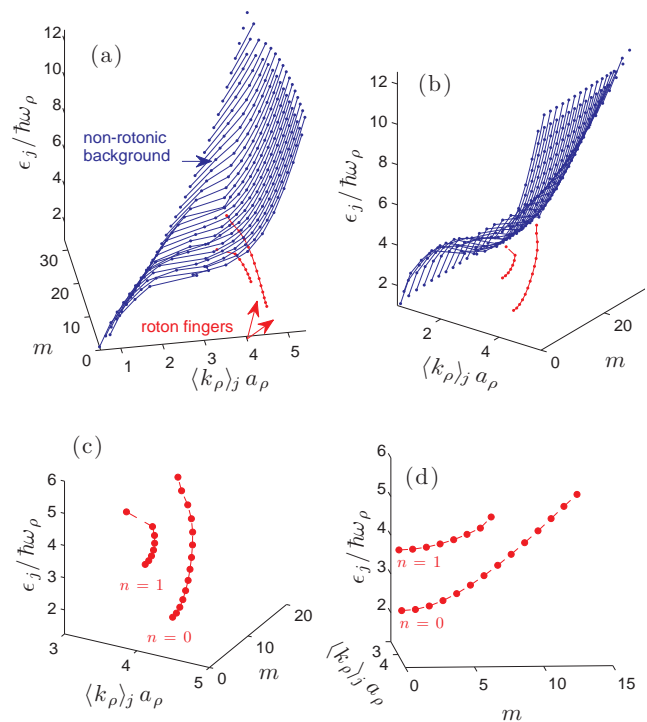} 
\caption{(color online) Roton fingers in the spectrum of a trapped dipolar BEC.  (a), (b) Two views of the quasiparticle excitations of a trapped dipolar condensate mapped against their angular momentum projection $m$ and effective linear momentum $\langle  k_{\rho}\rangle_j$ [see Eq.~(\ref{krho})]. The individual mapped excitations are represented by dots and separate into two categories: (i) a smooth \textit{non-rotonic background} part (blue) dots joined by lines to form a surface; (ii) the \textit{roton fingers} indicated by (red) dots that extend as linear chains below the non-rotonic background. (c), (d) Two views of the roton fingers for the same case shown in (a), (b).
Parameters: $\lambda=20$ and $D=220$. 
 \label{Fig:traprots}}
\end{center}
\end{figure}

In this paper we directly examine the structure and properties of the roton modes that emerge in a pancake shaped, trapped dipolar BEC using full three-dimensional numerical calculations. 
  We produce a dispersion relation-like characterization for the quasiparticle excitations  by mapping these excitations onto linear and  angular momentum, and use this to identify the rotons. Strikingly, in the trapped system the rotons emerge as fingers in the dispersion relation-like characterization (see Fig.~\ref{Fig:traprots}).    We then examine the properties of the rotons in each finger, as well as considering how the fingers change with the DDI strength.  Finally we show that by perturbing the harmonic trap with a repulsive Gaussian potential the  character of the roton fingers can be modified. Notably, we observe that modes with higher values of angular momentum projection become the minimum energy rotons in each finger, thus allowing a controllable way to produce \textit{angular rotons} (i.e.~where the lowest energy roton is one with a non-zero angular momentum projection) \cite{Ronen2007a}.

\section{Formalism}    

The condensate wavefunction (normalized to unity) is found by solving the non-local dipolar Gross-Pitaevskii equation (GPE) \cite{Goral2000a} for the lowest energy ground state
\begin{equation} 
\mu\psi_0 =\left[\!-\frac{\hbar^2\nabla^2}{2M}\!+\!V(\x)\!+\!\int d\x^\prime N_0U(\x\!-\!\x^\prime)| \psi_0(\x^\prime)|^2\right]\psi_0 ,
\end{equation}
where  $M$ is the atomic mass, $\mu$ is the chemical potential and  $N_0$ is the condensate number.  For dipoles  polarized along $z$  the inter-atomic interaction potential is of the form
\begin{equation} 
U(\mathbf{r}) =g\delta(\mathbf{r})+ \frac{3g_{\rm dd}}{4\pi}\frac{1-3\cos^2\theta}{|\mathbf{r}|^3},
\end{equation} 
where the short range interaction is characterized by the contact parameter $g=4\pi a_s\hbar^2/M$, with $a_s$ being the $s$-wave scattering length. The DDI parameter is   $g_{\mathrm{ dd}} = \mu_0\mu_m^2/3$, where   $\mu_m$ is the magnetic dipole moment, and $\theta$ is the angle between $\mathbf{r}$ and the $z$ axis.  In what follows we will consider a case where the contact interaction is tuned to zero (e.g.~by a Feshbach resonance). Results for non-zero contact interaction are qualitatively similar, and usually what is important is the proximity to the stability boundary (e.g.~see Fig.~4 and associated discussion of Ref.~\cite{Blakie2013a}). 

The atoms are taken to be confined by a harmonic potential 
\begin{equation}
V(\x) = \frac{1}{2}M\omega_\rho^2(\rho^2+\lambda^2z^2),
\end{equation} with  aspect ratio $\lambda=\omega_z/\omega_\rho$. Note that in Sec.~\ref{rotonengineering} we consider adding an additional perturbation to $V(\x)$.

The  fluctuations of the condensate are described by the field operator
 \begin{equation}
 \hat{\delta}(\mathbf{x}) \approx  \sum_j\left[u_j(\x)\hat\alpha_j - v_j^*(\x)\hat\alpha_j^\dagger\right],\label{psihatbog}
\end{equation}
where the quasiparticle modes $\{u_j,v_j\}$, with respective energies  $\epsilon_j$, are obtained by solving non-local Bogoliubov-de Gennes equations \cite{Ronen2006a}, which can also be obtained by linearizing about the Gross-Pitaevskii dynamics (e.g.~see \cite{Morgan1998a, Ronen2006a}).  The quasiparticle operators $\{\hat\alpha_j,\hat\alpha_j^\dagger\}$ satisfy standard bosonic commutation relations.
In pancake traps with $\lambda\gg1$ roton like excitations can emerge when the DDI is sufficiently strong, and is the regime we focus on here.   

 We adopt harmonic oscillator units defined by the radial trap frequency, i.e.~$\hbar\omega_{\rho}$  and $a_{\rho}=\sqrt{\hbar/M\omega_{\rho}}$ as the units of energy and length, respectively. We follow Ref.~\cite{Ronen2006a} and introduce  $D=3N_0g_{\rm dd}M/4\pi\hbar^2a_\rho$ as the dimensionless DDI parameter. To put this parameter into context of current experiments, the case of $D=220$, $\lambda=20$  (which we consider in Fig.~\ref{Fig:traprots})  corresponds to about $25\times10^3$ $^{164}$Dy atoms in a trap with $\omega_{\rho}=2\pi\times11$ s$^{-1}$.
Our numerical techniques for solving the dipolar GPE and Bogoliubov-de Gennes equations have been described elsewhere \cite{Blakie2013a}.
 
\section{Results}    
\subsection{Roton fingers}
We begin by considering the quasiparticle spectrum of a dipolar condensate in a $\lambda=20$ pancake shaped trap for DDI strength of $D=220$, which is sufficiently large for roton modes to develop. To visualize the excitations we plot each one  against its $z$ projection of angular momentum $m$ \footnote{Because the symmetry axis of the trap corresponds to the direction along which the dipoles are polarized, the system is invariant under rotations about the $z$ axis, and $m$ is a good quantum number.} and its effective linear momentum, which is assigned by  
\begin{equation} 
\langle k_{\rho}\rangle_j\equiv\sqrt{\frac{\int d\mathbf{k}\,k_{\rho}^2\left[| {\bar u}_j(\mathbf{k})|^2+|\bar v_j(\mathbf{k})|^2\right]}{\int d\mathbf{k}\, \left[| \bar{u}_j(\mathbf{k})|^2+|\bar v_j(\mathbf{k})|^2\right]} }, \label{krho}
\end{equation}
where $\bar{u}_j(\mathbf{k})=\mathcal{F}\{u_j(\mathbf{x})\}$ and $\bar{v}_j(\mathbf{k})=\mathcal{F}\{v_j(\mathbf{x})\}$ are the quasiparticle amplitudes in momentum space, with $\mathcal{F}$ the Fourier transform operator. We note that the mapping to an effective linear momentum for excitations in a trapped dipolar condensate was used in \cite{Wilson2010a} for $m=0,1,2$.

The results of this mapping, shown in Fig.~\ref{Fig:traprots}, provide a useful visualization of the quasiparticle excitations.  Figures \ref{Fig:traprots}(a) and (b) reveal that while the majority of modes form a reasonably smooth surface, two discrete \textit{fingers} of modes extend below this surface. These  fingers represent the roton modes in the trapped system. For clarity in what follows we will exclusively refer to modes within the fingers as \textit{rotons}. All remaining modes, constituting the smooth surface, will be referred to as the \textit{non-rotonic background} modes.

In Fig.~\ref{Fig:traprots}(c) and (d) we show the roton fingers in isolation and label the lower energy finger by the quantum number $n=0$, and the higher finger by $n=1$. As we show in Sec.~\ref{secrotoncharacter}, this quantum number corresponds to the number of nodes in the $k$-space wavefunctions of the quasiparticles.
Within the fingers the particular roton modes are then distinguished by their angular momentum projection $m$.
 Each finger has a minimum energy at $m=0$ and the energy of the finger modes increases with increasing $m$ up to some maximum (i.e.~$m=13$ for the $n=0$ finger, and $m=7$ for $n=1$) after which the finger joins the non-rotonic background \footnote{These general properties appear to hold while the condensate is in a \textit{normal state}, i.e.~has a smooth radial density profile with density maximum at trap center, which  monotonically decreases with increasing $\rho$ (cf.~Sec.~\ref{rotonengineering})}. The assignment of modes into roton fingers is unambiguous  except for the last (i.e.~highest $m$) modes which join up to the non-rotonic background. We find that this mapping procedure is applicable for geometries with $\lambda\gtrsim10$ where the fingers are sufficiently pronounced to distinguish from the non-rotonic background modes. The original results presented in Ref.~\cite{Ronen2007a} for trapped rotons considered cases with $\lambda\sim8$. We have applied our mapping to this case, and find that the fingers have $\sim2$ modes and are not easily distinguished from the non-rotonic background. In the cases we have examined we do not find any roton fingers developing in the dispersion relation-like branches for higher axial excitations (i.e.~$z$ excited modes).

\subsection{Roton mode character}\label{secrotoncharacter}

 \begin{figure}
\begin{center}
\includegraphics[width=3.3in]{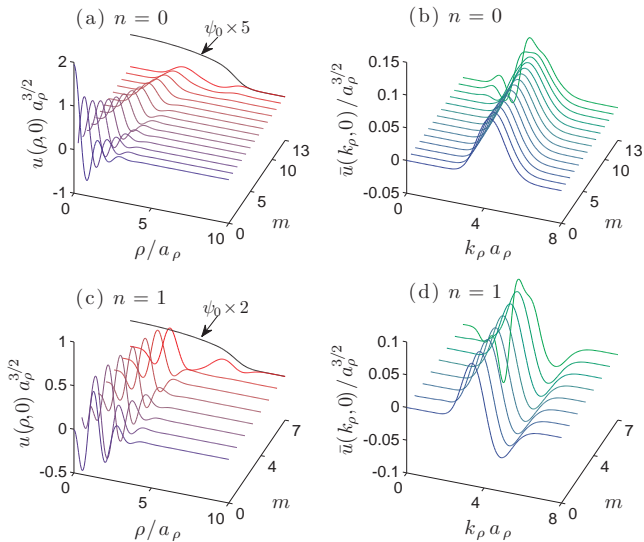} 
\caption{(color online) 
Roton mode functions in position and momentum space. The $u$ quasiparticle amplitudes for the roton modes are shown for the (a,b) $n=0$ and (c,d) $n=1$ fingers [as identified in Fig.~\ref{Fig:traprots}(d)] as a function of angular momentum projection. The rotons are shown in (a,c) position space along the $\rho$ axis, and in (b,d) momentum space along the $k_{\rho}$ axis.  The scaled condensate amplitude in position space is shown for reference in (a,c).
Note that the quasiparticle amplitudes are of the form $u(\mathbf{x})=u(\rho,z)e^{im\phi}$ etc., and we plot these along the $x$ axis, i.e.~taking $\phi=0$.
Parameters: $\lambda=20$ and $D=220$. 
 \label{Fig:rotuv}}
\end{center}
\end{figure}

\subsubsection{Roton mode functions in position and momentum space}
In Fig.~\ref{Fig:rotuv} we examine the radial  behavior of the $u$ quasiparticle amplitudes for the roton modes of both fingers. In position space [Fig.~\ref{Fig:rotuv}(a), (c)]  we see that the lowest $m$ modes in each finger are localized to the central region of the condensate while as $m$ increases the rotons begin to delocalize as the fingers join to the non-rotonic background. The momentum space [Fig.~\ref{Fig:rotuv}(b), (d)]  behavior of the roton modes shows that: (i) The modes are localized about $k_{\rho}\sim4/a_{\rho}\sim1/a_z$, where $a_z=\sqrt{\hbar/M\omega_z}$ is the $z$ confinement length (Note: $1/a_z\approx4.47/a_{\rho}$). This can be taken to define a roton wavevector $k_{\mathrm{rot}}$; (ii) The roton modes have a characteristic harmonic oscillator form (although displaced to be centered at  $k_{\mathrm{rot}}$), with the  $n$ quantum number corresponding to the number of nodes in the momentum wavefunction.

 \begin{figure}
\begin{center}
\includegraphics[width=3.4in]{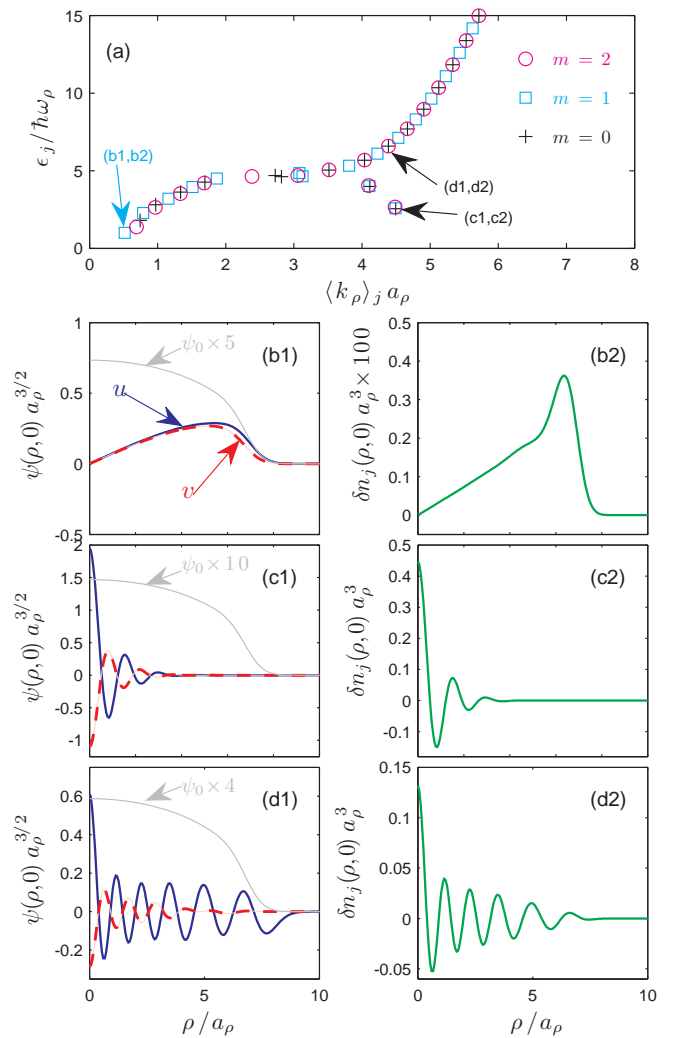} 
\caption{(color online)  Comparison of quasiparticle modes. (a) The spectrum of excitations mapped to a dispersion for the case in  Fig.~\ref{Fig:traprots}, only showing modes with angular momentum projection $|m|\le2$. The $m=1$ phonon, $m=0$ roton, and a non-rotonic background $m=0$ mode (of similar $\langle k_{\rho}\rangle_j$ to the roton mode) are identified with arrows. These modes are analyzed below in  subplots (b1,b2), (c1,c2) and (d1,d2), respectively.  (b1,c1,d1) Quasiparticle $u$  (solid blue line)  and $v$ (dashed red line) amplitudes, and the scaled condensate amplitudes (light gray line). (b2,c2,d2) Corresponding density fluctuations.
Parameters: $\lambda=20$ and $D=220$. 
 \label{Fig:phononvsrot}}
\end{center}
\end{figure}

\subsubsection{Phonon-Roton comparisons: density fluctuations}
In Fig.~\ref{Fig:phononvsrot} we compare a variety of quasiparticles including a phononlike mode, a roton mode, and a mode of similar effective linear momentum to the roton mode, but which resides in the non-rotonic background. In Fig.~\ref{Fig:phononvsrot}(a) we show a two-dimensional projection of the same data presented in Fig.~\ref{Fig:traprots}, but limited to $|m|\le2$, and identify the modes we use for this comparison. In addition to examining the behavior of the $u_j$ and $v_j$ amplitudes of the quasiparticles individually, it is of interest to consider the density fluctuations, $\delta \hat{n}\equiv \hat{n}-\langle\hat{n}\rangle$, where
$\hat{n}=N_0|\psi_0|^2+\sqrt{N_0}\psi_0(\hat{\delta}^\dagger+\hat{\delta})+\hat{\delta}^\dagger\hat{\delta},
$ 
is the density operator, and we have assumed that $\psi_0$ is real.
To leading order (taking $\hat\delta$ to be small) the density fluctuations are  given by
\begin{equation}
\delta \hat{n}(\mathbf{x})=\sqrt{N_0}\sum_j\left[\delta n_j(\mathbf{x})\hat{\alpha}_j+\delta n_j^*(\mathbf{x})\hat{\alpha}_j^\dagger\right],
\end{equation}
where $\delta n_j =\psi_0(u_j-v_j)$ is the density fluctuation amplitude associated with quasiparticle $j$.

 \begin{figure*}
\begin{center}
\includegraphics[width=6.5in]{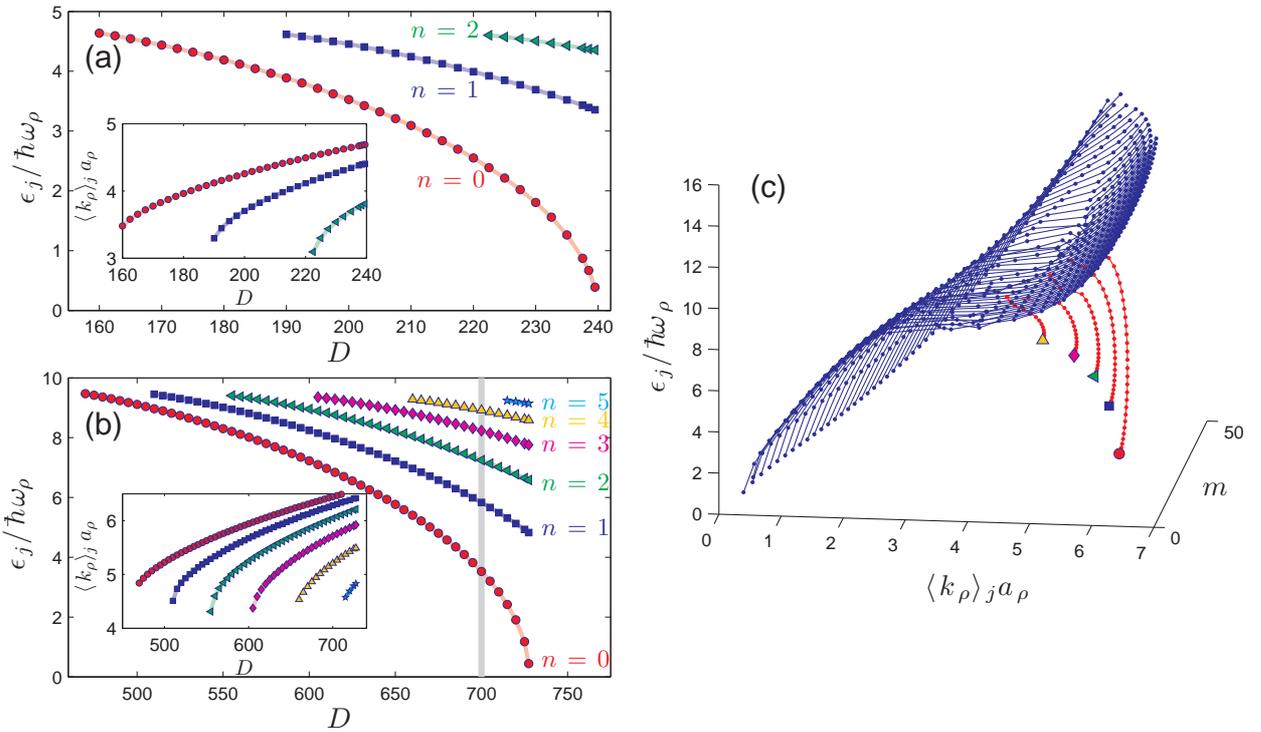} 
\caption{(color online)  Emergence and properties of roton fingers.  The energies of the lowest ($m=0$) roton in each finger for (a) $\lambda=20$, (b) $\lambda=40$.  Insets: the respective effective linear momenta of the $m=0$ rotons in each finger.  (c) The dispersion relation-like characterization of a $\lambda=40$ dipolar BEC at $D=700$ [this case is indicated by a gray vertical line in (b)].  The $m=0$ rotons at the start of each finger are marked with the symbols used to represent them in (b). 
 \label{Fig:rotspec}}
\end{center}
\end{figure*}

 The $u_j$ and $v_j$ components of the phonon mode [Fig.~\ref{Fig:phononvsrot}(b1)] extend over the size of the condensate. Since the $u_j$ and $v_j$  are in-phase and almost equal, the associated density fluctuation   $\delta n_j$ is small  [Fig.~\ref{Fig:phononvsrot}(b2)]. The roton mode [Fig.~\ref{Fig:phononvsrot}(c1)]  has a short wavelength and is localized  near the center of the condensate. Due to this localization, and because the  $u_j$ and $v_j$ amplitudes of the rotons are out-of-phase, the associated density fluctuation has a peak value that is approximately 100 times larger than for the phonon mode. This difference in behavior occurs because of the momentum dependence of the interaction: the interaction between the condensate and its excitations is repulsive (i.e.~suppressing density fluctuations) at long wavelengths, whereas it becomes attractive (i.e.~enhancing density fluctuations) at wavelengths similar to $a_z$ (also see Fig.~5 of Ref.~\cite{Blakie2013a}). 
The enhanced density fluctuations of dipolar BECs has also  been identified in Refs.~\cite{Klawunn2011,Boudjemaa2013a,Bisset2013a}. For comparison a mode with similar effective linear momentum to the roton mode, but not within the roton finger, is shown in  Fig.~\ref{Fig:phononvsrot}(d1).

\subsubsection{Development of fingers}\label{Sec:DevelopFingers}
How the fingers emerge as DDI strength $D$ increases is considered in Fig.~\ref{Fig:rotspec}, where we present results for both $\lambda=20$ and $\lambda=40$. The first fingers appear when $D$ is sufficiently large (for $\lambda=20$ the fingers first emerge at $D\sim160$ and for $\lambda=40$ they emerge at $D\sim450$). As $D$ increases the fingers decrease in energy and become longer (i.e.~extend to over a larger $m$ range), and additional fingers emerge from the non-rotonic background. 
 For  a sufficiently large DDI strength $D_{\mathrm{crit}}$  the lowest ($n=0$) finger will fall to zero energy \footnote{Usually the $m=0$ mode in the $n=0$ finger goes soft, but in special parameter regimes the condensate can take a bi-concave shape and a mode with $m\ne0$  softens \cite{Ronen2007a} (also see Sec.~\ref{rotonengineering})},  and  will become imaginary for $D>D_{\mathrm{crit}}$, signaling that the condensate is dynamically unstable \cite{Ronen2006a,Ronen2007a}. For higher trap aspect ratios a greater number of fingers emerge before the onset of the dynamical instability, e.g.~for $\lambda=20$  we find that 3 fingers emerge by $D_{\mathrm{crit}}\approx240$  [see Fig.~\ref{Fig:rotspec}(a)];
for $\lambda=40$  we find that 6 fingers emerge by $D_{\mathrm{crit}}\approx728$  [see Fig.~\ref{Fig:rotspec}(b)].

 \begin{figure} 
\begin{center}
\includegraphics[width=3.3in]{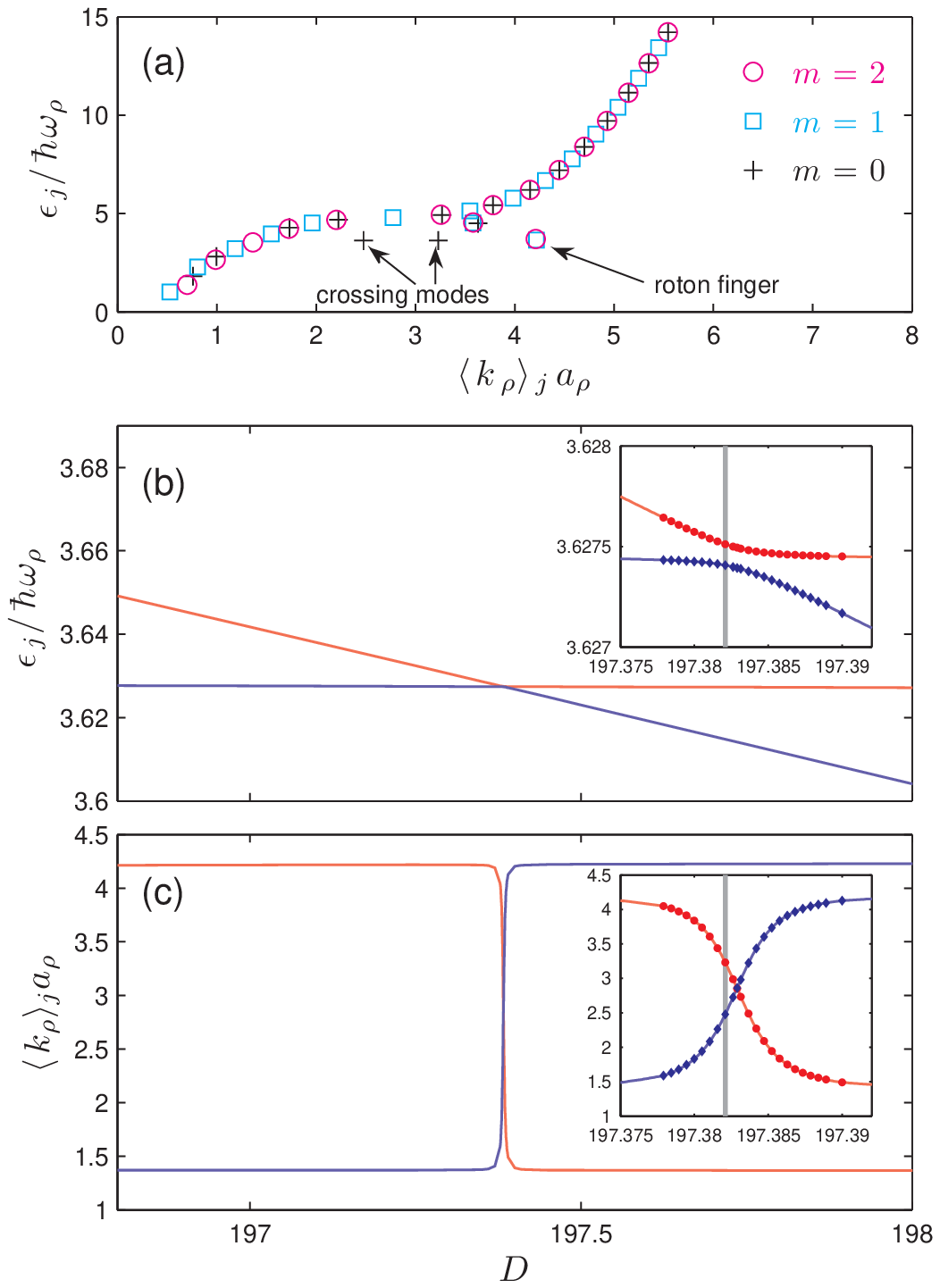} 
\caption{(color online)  
Avoided crossings of phonon (typically low $\langle k_\rho\rangle_j$) and roton (typically high $\langle k_\rho\rangle_j$) modes. (a) The spectrum of excitations mapped to a dispersion for $D=197.382$ only showing modes with angular momentum projection $|m|\le2$. The $m=0$ phonon and roton modes,  which exhibit the avoided crossing,  are identified with arrows, as is the location of the finger that the dislocated mode originated from.   The values of the (b)  quasiparticle energies and (c) effective linear momentum $\langle k_{\rho}\rangle_j$  of the two modes identified in (a) as the DDI strength is varied. Insets to (b) and (c) show an enlarged view of the data near the avoided crossing. Gray vertical lines mark the case considered in (a). Note, trap aspect ratio is $\lambda=20$. 
 \label{Fig:rotavoid}}
\end{center}
\end{figure}

\subsubsection{Finger dislocations: Phonon-Roton avoided crossings}
We also note that while the roton fingers are generally smooth functions of $m$ [see Figs.~\ref{Fig:traprots}(d), \ref{Fig:rotspec}(c)], for certain parameters we observe that particular roton modes dislocate from the finger by having a $\langle k_{\rho}\rangle_j$ value that is significantly less than the other modes in that finger. The origin of these dislocations is avoided crossings between roton and phonon modes in the same $m$ subspace \footnote{In our cylindrically symmetric treatment only modes in the same $m$ subspace can couple to each other.}. 
We demonstrate this in Fig.~\ref{Fig:rotavoid} where we consider such a crossing that affects the $n=0,m=0$ roton.
In Fig.~\ref{Fig:rotavoid}(a) we show the roton mode dislocated from the finger in the midst of such a crossing, noting that the coupled phonon mode undergoes a matching dislocation to a higher momentum value. 

To explore this crossing we vary the DDI strength. As $D$ increases the roton mode energy decreases, \textit{crossing} the relevant phonon mode energy  [see Fig.~\ref{Fig:rotavoid}(b)]. Due to the coupling between these modes they undergo an avoided crossing during which the two modes hybridize, leading to a significant change in  $\langle k_{\rho}\rangle_j$ [Fig.~\ref{Fig:rotavoid}(c)].  We emphasize that these avoided crossings can occur for any value of $m$, however because the coupling between phonon and roton modes is weak they tend to occur in very narrow parameter regimes.

\subsection{Relation to predictions of Jona-Lasinio \textit{et al.} \cite{JonaLasinio2013}}
In Ref.~\cite{JonaLasinio2013} Jona-Lasinio \textit{et al.}~developed an analytic description of rotons in a trapped dipolar BEC. We briefly review their results and comment on its relationship to our full numerical treatment

A central idea of Ref.~\cite{JonaLasinio2013} is that after integrating out the tightly confined ($z$) degree of freedom, a local quasiparticle spectrum $\epsilon(k_{\rho},\rho)$ can be obtained for the in-plane coordinates, with the $\rho$ dependence arising from the variation of the condensate density.  When the system exhibits a roton it emerges as a local minimum in $\epsilon(k_{\rho},\rho)$ at finite wavevector $k_{\rho}=k_{\mathrm{rot}}$ and at the trap center where the condensate density is highest. Expanding about this minimum to second order yields
\begin{align}
\epsilon(k_{\rho},\rho)&\approx \epsilon_{\min}+\frac{\hbar^2}{2M_*}(k_{\rho}-k_{\mathrm{rot}})^2+\frac{1}{2}m_*\omega_*^2\rho^2,\label{eLDA}
\end{align}
where the energy minimum $\epsilon_{\min}$, the roton effective mass $M_*$ and the effective confinement frequency $\omega_*$ have been introduced.  By re-quantizing Eq.~(\ref{eLDA}) in momentum space, a spectrum of the form 
\begin{equation}
E_{nm}\simeq\left[\frac{m^2-\tfrac{1}{4}}{2(k_{\mathrm{rot}}l_*)^2}+n+\frac{1}{2}\right]\hbar\omega_*,\label{Enm}
\end{equation}
is obtained, valid for  $k_{\mathrm{rot}}l_*\gg1$, where $l_*=\sqrt{\hbar/M_*\omega_*}$.

Equation (\ref{Enm}) should approximately describe the spectrum of our roton fingers. The physics used as the basis of this approach indeed qualitatively  explains many of the results we observe, e.g.~the localization in position and momentum space, and the harmonic oscillator like form of the roton wavefunctions in momentum space [see Fig.~\ref{Fig:rotuv}]. Our roton spectrum does exhibit a quadratic dependence on $m$ for small values of $m$ [e.g.~see Fig.~\ref{Fig:traprots}(d)], as predicted in Eq.~(\ref{Enm}), but this eventually crosses over to being approximately linear in $m$ for higher $m$ values. 
More generally, within the parameter regime we have studied (i.e.~$\lambda=20$ and $\lambda=40$ pure dipolar BECs) we do not obtain quantitative agreement with the analytic spectrum. This can be seen from Fig.~\ref{Fig:rotspec} where the fingers are not equally spaced in energy, as predicted by Eq.~(\ref{Enm}). This suggests that anharmonic corrections to  Eq.~(\ref{eLDA}) are significant. This anharmonicity is also evident because the higher energy modes (i.e.~both higher $m$ and $n$ values) tend to shift to lower values of effective linear momentum [see insets to Figs.~\ref{Fig:rotspec}(a) and (b), and Figs.~\ref{Fig:traprots}(c) and \ref{Fig:rotspec}(c)]. This means that we cannot unambiguously assign a single $k_{\mathrm{rot}}$ to describe the rotons in the trapped system.
Many of the anharmonic effects may be described by the full local quasiparticle spectrum introduced in \cite{JonaLasinio2013} prior to making the quadratic approximation [i.e.~Eq.~\ref{eLDA}] \footnote{Private communication L.~Santos}.

\subsection{Engineering the roton spectrum}\label{rotonengineering}
 
In this section we consider the effect that a more general type of external potential has on the rotons in a dipolar BEC. To do this we add to our harmonic trap a Gaussian perturbing potential  of the form
\begin{equation}
V_{\mathrm{pert}}(\x) = V_0\exp\left(-\frac{\rho^2}{\sigma_0^2}\right),
\end{equation}
characterized by its strength $V_0$  and width $\sigma_0$.  With this addition the combined potential is flat bottomed trap for small  positive values of $V_0$ ($V_0<\tfrac{1}{2}M\omega_{\rho}^2\sigma_0^2$ \cite{Baillie2010a}) and becomes toroidal shaped for larger $V_0$.  We restrict our attention to perturbing potentials with $\sigma_0$ much greater than the roton wavelength $\sim a_z$ to ensure that the added potential smoothly modifies the condensate density profile rather than coupling strongly to the rotons and giving rise to strong density oscillations (e.g.~see \cite{Yi2006a,Wilson2008a}). We note that the effect of a barrier type perturbation in the $z$ direction was considered in Ref.~\cite{Asad-uz-Zaman2010a}, and found to enrich the stability properties compared to the harmonically trapped case.  Also Lu \textit{et al.}~\cite{Lu2010a} examined the properties of a condensate in a  box potential, and predict the formation of spatial density oscillations in the condensate due to the sharp  edges of the potential.

 \begin{figure} 
\begin{center}
\includegraphics[width=3.4in]{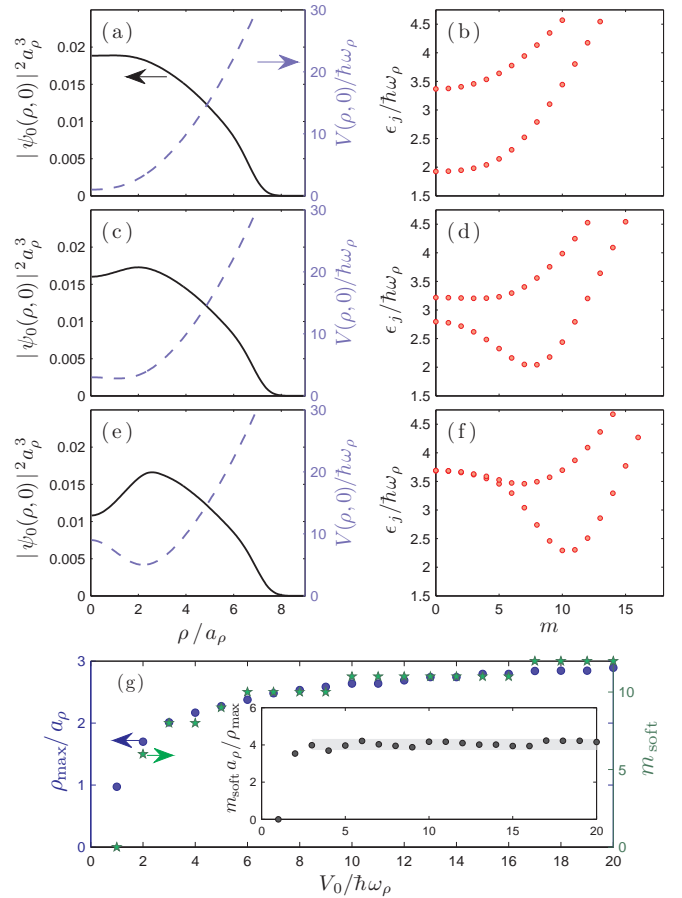} 
\caption{(color online)  Condensate density profile and roton finger structure under the influence of a perturbing potential.
(a,c,e) The condensate density profile (solid black line) and the total external potential (dashed blue line) for the various cases considered. (b,d,f)  Lowest two roton fingers obtained from the corresponding quasiparticles. (g) The radial location of the condensate maximum density $\rho_{\max}$ and the $m$ value of the lowest energy roton $m_{\mathrm{soft}}$ as $V_0$ increases. Inset: $m_{\mathrm{soft}}/\rho_{\max}$ ratio for the data in the main plot. Shaded thick line shows that as $V_0$ increases the ratio remains approximately constant. Arrows indicate the relevant vertical axis for data in subplots.
Results for  $\sigma_0=2\,a_{\rho}$  and (a,b) $D=251.26$, $V_0=1\,\hbar\omega_{\rho}$; 
(c,d) $D=274.47$, $V_0=3\,\hbar\omega_{\rho}$
 (e,f) $D=285.53$,   $V_0=9\,\hbar\omega_{\rho}$. Values of $D$ for all results  chosen  so that $D \max\{|\psi_0|^2\}a_\rho^3=4.740$, to ensure the system is approximately the same distance from the stability boundary. \label{Fig:toroid}}
\end{center}
\end{figure}

In Fig.~\ref{Fig:toroid} we show the condensate density and the roton fingers for a perturbation of width $\sigma_0=2\,a_{\rho}$. For $V_0=1\,\hbar\omega_{\rho}$ [Fig.~\ref{Fig:toroid}(a)] the perturbing potential flattens the central  condensate density, and causes the energies of the first few ($m\lesssim3$) $n=0$ modes to become almost degenerate [Fig.~\ref{Fig:toroid}(b)].
As the strength of the perturbing  potential increases the condensate develops a local minimum in density at trap center [Figs.~\ref{Fig:toroid}(c), (e)]. When this occurs the roton fingers take a different character: the minimum energy in each roton finger occurs at a non-zero angular momentum projection ($m_{\mathrm{soft}}$) [Figs.~\ref{Fig:toroid}(d), (f)].  In these cases the lowest energy roton propagates around the ring of maximum condensate density of radius $\rho_{\max}$, causing azimuthal fluctuations in the density of this ring. We also note that when the condensate density is sufficiently bi-concave [as in Fig.~\ref{Fig:toroid}(e)], the first two fingers of the roton spectrum become degenerate, having the same energy and $\langle k_{\rho}\rangle$ values,  at low   $m$ [as in Fig.~\ref{Fig:toroid}(f)].  This appears to occur when the roton modes become well localized (spatially) in the dense condensate ring, and have a small amplitude at trap center.

We find that the $m$ value of the softest   roton mode ($m_{\mathrm{soft}}$) is related to the radius at which the condensate density maximum occurs, and can be controlled by adjusting the height and width of the perturbation potential.   Assuming that the softest mode causes an azimuthal buckling around the condensate ring at the roton wavelength we expect that $k_{\mathrm{rot}}\approx m_{\mathrm{soft}}/\rho_{\max}$. From this it follows that  $m_{\mathrm{soft}}$ should increase proportional to  $\rho_{\max}$ if $k_{\mathrm{rot}}$ is assumed to be constant.   This relationship is verified in Fig.~\ref{Fig:toroid}(g) and inset.

\section{Conclusion and outlook} 
In this paper we have explored the properties of  rotons in pancake shaped dipolar BECs using fully three-dimensional solutions of the dipolar Gross-Pitaevskii and Bogoliubov-de Gennes equations. We have shown that rotons in the trapped system emerge as finger like chains when visualized using a momentum mapping technique, and that the quasiparticle modes are localized in both position and momentum space.  Ronen \textit{et al.}~\cite{Ronen2007a}  established the connection between the condensate developing a modulated (bi-concave) density profile and the condensate becoming dynamically unstable due to a so called \textit{angular roton} (i.e.~a roton mode with $m\ne0$). They studied spontaneously occurring bi-concave states, i.e.~those which occur in a pure harmonic trap in particular regions of interaction and trap geometry parameter space.
For large aspect ratio traps, where the roton modes become clearly identifiable, such spontaneous  bi-concave states only occur near the stability boundary  (e.g.~see \cite{Lu2010a,Blakie2012a}), and may be difficult to access experimentally.  
Importantly, our results show that  imposing  the density modulation using a perturbation also gives rise to  angular rotons, and that these can be produced reasonably far from the stability boundary. Furthermore, by adjusting the perturbation, angular rotons with large angular momentum projection can be controllably engineered.

Important questions remain about the regime in which meanfield predictions may be quantitatively accurate. For the uniform dipolar BEC it has been shown that density fluctuations arising from the rotons can be significant at finite temperature  \cite{Boudjemaa2013a} (also see \cite{Ronen2007b,*Lima2011a,*Ticknor2012a,*Bisset2012,*Lima2012a,*Pawlowski2013}). Thus Bogoliubov theory may be limited to low temperatures $T\ll T_c$, where $T_c$ is the condensation temperature \cite{Glaum2007}.

Finally, we briefly discuss how aspects of rotons might be verified in experiments. One possibility is to measure the number fluctuations within finite-sized cells, as has been implemented in a number of experiments with pancake shaped condensates (e.g.~see \cite{Hung2011a}). It was shown in Ref.~\cite{Bisset2013a} that such measurements can be made sensitive to individual $m=0$ roton modes, revealing the location and size of the roton mode. Another possibility suggested in Ref.~\cite{Wilson2009a} is that the shape of the lowest energy mode can be revealed by quenching the system across the stability boundary (e.g.~by  making the $s$-wave scattering length negative). In particular, this procedure reveals the presence of an angular roton through the development of a nontrivial angular distribution in the system post quench (c.f.~the $d$-wave symmetry of a dipolar BEC collapsing from a nearly spherical trap observed in Ref.~\cite{Lahaye2009a}). An alternative approach is to use some form of energy sensitive spectroscopy technique, such as collective mode or Bragg spectroscopy, which have already seen initial applications to dipolar BECs  \cite{Bismut2010a,Bismut2012a}.

  \section*{Acknowledgments} We acknowledge valuable feedback on the manuscript from L.~Santos.  This work was supported by the Marsden Fund of New Zealand (contracts UOO0924 and UOO1220). RB acknowledges support from the Graduate Research Committee, by means of the UO   Publishing Bursary (Doctoral).

\bibliographystyle{apsrev4-1}
 
%

\end{document}